\documentclass[
    11pt,
    letterpaper,
    reprint,
    notitlepage,
    superscriptaddress,
    aps,prx
]{revtex4-2}

\usepackage{amsmath,amssymb,amsfonts} 
\usepackage{empheq}
\usepackage{physics}
\usepackage{lipsum}

\usepackage[normalem]{ulem}

\usepackage{microtype}

\usepackage[svgnames,dvipsnames]{xcolor}
\usepackage{graphicx}
\definecolor{NewBlue}{rgb}{0.1, 0.1, 0.7}
\definecolor{NewRed}{rgb}{0.7, 0.1, 0.1}
\usepackage[colorlinks,
    linkcolor=Maroon,
    citecolor=NewBlue,
    urlcolor=ForestGreen]{hyperref}

\usepackage[capitalise]{cleveref}

\usepackage[
    exponent-product=\cdot,
    range-phrase=\text{--},
    range-units=single,
    separate-uncertainty,
]{siunitx}[=v2]

\renewcommand{\t}[1]{\mathrm{{#1}}}

\newcommand{\dop}[1]{\delta\hat{#1}}
\newcommand{\avg}{\expval}

\usepackage{empheq}

\newcommand{\LigoMIT}{LIGO Laboratory, Massachusetts Institute of Technology, Cambridge, MA 02139}
\newcommand{\MechMIT}{Department of Mechanical Engineering, Massachusetts Institute of Technology, 
    Cambridge, MA 02139}
\newcommand{\Utah}{Department of Electrical and Computer Engineering, University of Utah, Salt Lake City, Utah 84112}

\begin{document}

\title{Active laser cooling of a centimeter-scale torsional oscillator}

\author{Dong-Chel Shin}
\email{dongchel@mit.edu}
\affiliation{\MechMIT}

\author{Tina Hayward}
\affiliation{\Utah}

\author{Dylan Fife}
\affiliation{\MechMIT}

\author{Rajesh Menon}
\affiliation{\Utah}

\author{Vivishek Sudhir}
\email{vivishek@mit.edu}
\affiliation{\MechMIT}
\affiliation{\LigoMIT}

\date{\today}

\begin{abstract}
Experimental tests of gravity’s fundamental nature call for mechanical systems in the quantum regime while being sensitive to gravity.
Torsion pendula, historically vital in studies of classical gravity, are ideal for extending gravitational tests into the quantum realm due to their inherently high mechanical quality factor, even when mass-loaded.
Here, we demonstrate laser cooling of a centimeter-scale torsional oscillator to a temperature of 10 mK 
(average occupancy of 6000 phonons) starting from room temperature. 
This is achieved by optical radiation
pressure forces conditioned on a quantum-noise-limited optical measurement of the torsional mode
with an imprecision 9.8 dB below its peak zero-point motion.
The measurement sensitivity is the result of a novel `mirrored' optical lever that passively rejects extraneous 
spatial-mode noise by 60 dB.
The high mechanical quality ($1.4\times 10^7$) and quantum-noise-limited measurement imprecision demonstrate the
necessary ingredients for realizing the quantum ground state of torsional motion --- a pre-requisite for 
mechanical tests of gravity's alleged quantum nature.
\end{abstract}

\maketitle

\section{Introduction}
Torsion pendula have long been pivotal in the measurement of weak fundamental forces, most notably 
in establishing the electrostatic inverse-square law \cite{Coul,GillRitt93},
precision measurements of classical gravitational forces \cite{Cav,Adel03,Adel09,SpeQui14,RothSch17}, 
tests of the equivalence principle \cite{Dicke61,BragPan72,SchAdel08,wagner_torsion-balance_2012}, 
and the first observation of radiation pressure torque \cite{Beth36}.
In all these experiments, the torsion pendulum is employed as a sensor for a weak classical force. 

Recent interest in observing gravity's alleged quantum nature calls for
experiments where gravitationally attracting macroscopic mechanical oscillators are simultaneously prepared in
quantum states of their motion \cite{DattMia21,LamPlen24,KrySud23}. Torsion pendula are particularly suited for
such experiments on account of the low thermal Brownian noise of torsional suspensions even 
when mass-loaded \cite{Quinn97,PraWil23}, and well-understood techniques for isolating gravitational interaction
between them even with masses as small as 100 mg \cite{WestAsp21}.
However, in contrast to the mature array of techniques available for quantum-limited measurement and control of 
linear motion within the field of cavity optomechanics \cite{Aspelmeyer2014},
experimental realization of similar techniques has remained
elusive for torsional motion.
While levitated optomechanics has achieved significant progress \cite{GonRom21}, including the quantum measurement, control, and ground-state cooling of both linear and librational motions \cite{vander2021,ahn2018,kamba2022,kuhn2017,pontin2023}, these techniques remain constrained to nanoscopic scales.

In this letter, we demonstrate laser cooling of a centimeter-scale high-quality torsional oscillator using a 
novel ``mirrored optical lever'' whose quantum-noise-limited sensitivity of $10^{-12}\, \mathrm{rad/\sqrt{Hz}}$ 
is 9.8 dB below the peak zero-point motion of the torsional mode. 
Conditioned on this measurement, we apply optical radiation 
pressure torque on the oscillator so as to cool its angular motion to ${10}\, \mathrm{mK}$ from room temperature, corresponding to an average phonon occupation of $5964\pm39$ (starting from $\sim 2\times10^8$).
In the following, we describe the torsional oscillator, the mirrored optical lever used to measure its angular motion,
its performance in terms of classical noise cancellation, its calibration, and utility in measurement-based 
feedback cooling of torsional motion.

\begin{figure*}[t!]
    \includegraphics[width=2\columnwidth]{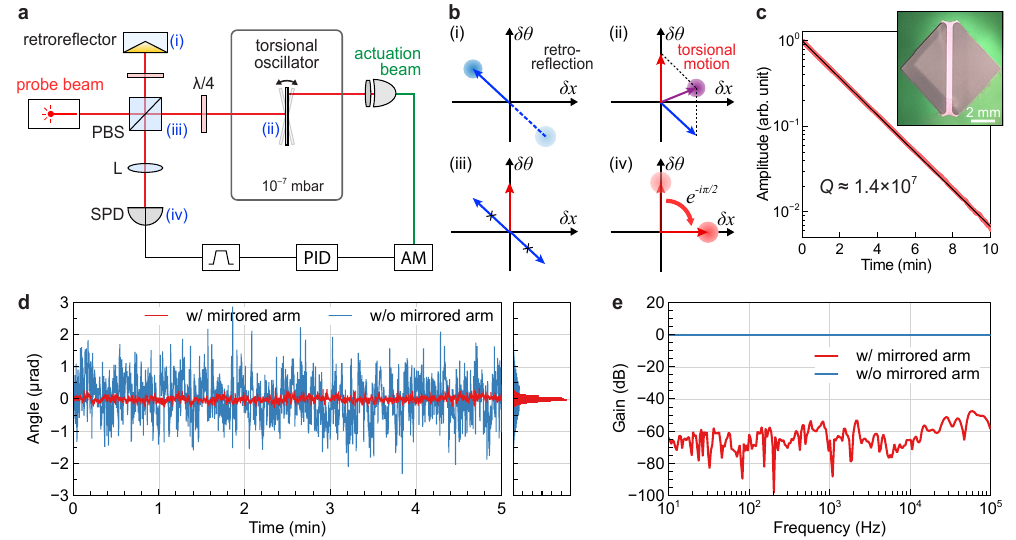}
    \centering
    \caption{\label{fig:concept}
    (a) Experimental setup: 1064 nm light in the fundamental spatial mode is used to measure and actuate a cm-scale torsional oscillator (panel (c)).
    The probe beam is split into two paths: one arm to the torsional oscillator (signal arm), and the other to a retroreflector (reference arm).
    The fields from the two arms are collected and subsequently detected by the split photodetector (SPD).
    Here $\lambda/4$: quarter-wave plate; L: lens; PBS: polarizing beam-splitter.
    (b) Principle of spatial mode noise suppression in the mirrored optical lever.
    The retroreflector ((i)) produces a mirror-image of the laser beam's spatial fluctuations (blue arrow), while the torsional oscillator ((ii)) induces a pure tilt motion (red arrow) in the signal arm.
    At the SPD ((iv)), the classical spatial mode noise of the laser beam is canceled.
    (c) A cm-scale torsional oscillator fabricated in tensile-stressed SiN (inset) features a mechanical quality factor of $10^7$ due to torsional dissipation dilution. 
    (d) Ambient angle fluctuation over time, measured with and without the mirrored arm.
    (e) Magnitude response of the mirrored optical lever: a sinusoidal tilt modulation is generated by the acousto-optic deflector and detected by lock-in amplification of the the split photodetector signal. The magnitude of this quantity,
    with and without the mirrored arm, characterizes the suppression of input classical tilt noise, here at the level 
    of 60 dB.
    }
\end{figure*}

\section{High-Q centimeter-scale torsional oscillator}
A remarkable advantage of torsional pendula in studies of gravity is that their suspensions can be realized with 
exceptionally high mechanical quality factor ($Q$) even when mass-loaded \cite{Quinn97,PraWil23}. 
The reason is two-fold \cite{PraWil23}: in a doubly-clamped bifilar (or ribbon-shaped) torsional suspension, 
tensile stress leads to dilution of the intrinsic mechanical dissipation, and the bifilar geometry 
is naturally soft-clamped so that loss at the clamps, and in the suspended mass, is suppressed.
Thus the intrinsic quality factor, \(Q_{\mathrm{int}}\), is elevated to \cite{PraWil23}
$Q_0 = Q_\mathrm{int}D_Q$ where the dissipation dilution factor is $D_Q \approx (\sigma/2E)(w/h)^2$;
here \(E\) and \(\sigma\) are Young's modulus and tensile stress, and \(w\) and \(h\) are the width and 
thickness of the ribbon. 
This is in marked contrast to tensile-stressed mass-loaded linear oscillators, where bending curvature due to the
loaded mass undermines the advantage from dissipation dilution \cite{ShanReg23,LiuSil21}.
This capability positions a macroscopic high-Q torsion pendulum as a unique candidate for both reaching 
its motional ground state and for gravitational experiments.

We fabricated a doubly-clamped 0.9-cm long thin-film ($w=\SI{0.5}{mm}$, $h=\SI{400}{nm}$) 
tensile-stressed ($\sigma = \SI{0.8}{GPa}$) torsional oscillator made of
stoichiometric $\mathrm{Si_3 N_4}$.
The device was fabricated starting from a double-sided $\mathrm{Si_3N_4}$-on-Si 
wafer followed by lithography and reactive ion
etching. A second aligned lithography and etch created the window for optical access from the back-side. 
The device was released using a 24-hour etch in potassium hydroxide. The sample underwent meticulous cleaning with acetone, isopropyl alcohol, deionized water, and oxygen plasma (see SI for more details).
The specific device used in the current study has its fundamental torsional mode resonating at
$\Omega_0 = 2\pi\cdot \SI{35.95}{kHz}$ with quality factor $Q = 1.4\times 10^7 $ inferred by ringdown measurements 
(in vacuum, at $\SI{6e-7}{mbar}$) as shown in \cref{fig:concept}(c).
The measured $Q$ is consistent with the expected dilution factor $D_Q \approx 2300$.
The moment of inertia of the fundamental torsional mode is measured to be $I=\SI{5.54}{kg\cdot m^2}$ (see SI), with an effective mass $m=I(2/w)^2=\SI{0.89}{\mu g}$ \cite{PraWil23}. 
Importantly, the design principles demonstrated here ensure that when a mass is added, the quality factor will not degrade \cite{manley2024}.

\section{Quantum-limited optical lever detection}

To achieve quantum-limited readout of angular motion, 
we devised a `mirrored' optical lever. Its primary
advantage is passive rejection of classical noises arising from the laser beam's transverse displacement and tilt.
Suppose the output of a laser is predominantly in the fundamental Hermite-Gaussian mode with 
amplitude $\bar{a} = \sqrt{P/\hbar \omega_\ell}$ ($P$ the optical power, $\omega_\ell = 
2\pi c/\lambda$ the carrier frequency for a wavelength $\lambda \approx \SI{1064}{nm}$), then the optical field
can be expressed as \cite{EnoKawa16,HaoPur24} $\hat{E}_\mathrm{in}(\vb{r},t) = (\bar{a} + \delta\hat{a}_{00}(t)) U_{00}(\vb{r}) 
+ \sum_{n,m} \delta\hat{a}_{nm}(t) U_{nm}(\vb{r})$. Here $U_{nm}(\vb{r})$ is the $(n,m)-$Hermite-Gaussian ($\mathrm{HG}_{nm}$)
basis function (see SI); the operators $\{\delta\hat{a}_{nm}(t)\}$ represent
fluctuations in the laser field, which in the ideal case, when the field is quantum-noise-limited, 
model the quantum vacuum fluctuations in $\mathrm{HG}_{nm}$ mode.
If this incident field is subjected to a transverse displacement $\delta \vb{r} = (\delta x,0,0)$ and angular tilt
$\delta \theta $ at its beam waist, the optical field is transformed to 
$\hat{E}_\mathrm{out}(\vb{r},t) \approx \hat{E}_\mathrm{in}(\vb{r},t) + \bar{a}(\delta x/w_0 + 
ikw_0\delta\theta/2)U_{10}(\vb{r})$,
where $w_0$ is the waist size. 
That is, transverse or angular motion scatters light from the incident $\mathrm{HG}_{00}$ mode to the 
$\mathrm{HG}_{10}$ mode in proportion to the motion.
When the beam waist is placed at the location of the torsional oscillator, its physical motion $\delta\hat{\theta}_\mathrm{phys}(t)$ 
is experienced twice by the reflected optical beam, i.e. $2\delta\hat{\theta}_{\mathrm{phys}}(t)$; 
the incident beam may also have extraneous classical 
noises in the transverse displacement ($\delta\hat{x}_{\mathrm{ext}}$) and tilt ($\delta\hat{\theta}_{\mathrm{ext}}$). 
When the reflected beam is detected by a split balanced photodetector (SPD) downstream, the resulting
photocurrent fluctuations are described by its (symmetrized single-sided) power spectral density 
(see SI, we neglect correlations between the tilt and transverse displacement \cite{HaoPur24})
\begin{equation}
    \begin{split}
    S_{I}[\Omega] = \frac{2}{\pi}(2\eta R & P k w_0 \sin{\zeta})^2
    \Big[S_{\theta}^{\mathrm{phys}}[\Omega] + \frac{1}{4}S_{\theta}^{\mathrm{ext}}[\Omega]\\
    & + \left( \frac{\cot{\zeta}}{k w_0^2} \right)^2 S_{x}^{\mathrm{ext}}[\Omega] 
    + \frac{\pi/2}{2\eta (\bar{a}k w_0)^2}\csc^2{\zeta}\Big], \label{PSD1}
    \end{split}
\end{equation}
where $S_{\theta}^{\mathrm{phys}}=S_{\theta}^{\mathrm{int}}+S_{\theta}^{\mathrm{ba}}$ represents the angular displacement spectral density of the torsional oscillator, which includes the contributions from the intrinsic motion $S_{\theta}^{\mathrm{int}}$ and the quantum back-action $S_{\theta}^{\mathrm{ba}}$. Additionally, $R$ is the responsivity of the detector, $\eta$ is the detection efficiency, 
and $\zeta$ is the Gouy phase shift from the torsional oscillator to the SPD.
It is worth noting that the Gouy phase shift indicates the degree to which the angular tilt of the beam is converted into transverse displacement during propagation, enabling the SPD to detect the beam tilt angle.
The last three terms represent the imprecision in the measurement arising from extraneous tilt noise
($S_\theta^\mathrm{ext}$), extraneous transverse displacement noise ($S_x^\mathrm{ext}$), and the fundamental 
imprecision from quantum vacuum fluctuations (from all odd-order HG modes that the SPD is sensitive to; 
see SI).
Thus, quantum-noise-limited detection of torsional motion is only possible if extraneous 
noise in the spatial mode of the laser beam is suppressed.

Our mirrored optical lever technique (shown in \cref{fig:concept}a,b) cancels classical spatial mode noise 
to achieve quantum-noise-limited detection. 
Specifically, the laser beam is split by a polarizing beam splitter (PBS) into two paths: one arm to the torsional oscillator (signal arm), and the other to a retroreflector (reference arm). 
A corner cube retroreflector produces a mirror-image of the laser beam's transverse displacement and tilt, 
i.e. $\delta\hat{a}_{10} \rightarrow -\delta\hat{a}_{10}$ after interacting with the retroreflector 
(\cref{fig:concept}(b)(i) shows the effect of this on input noise), whereas the torsional oscillator induces 
pure tilt (\cref{fig:concept}(b)(ii)). 
The fields from the two arms are collected by quarter-wave plates at the PBS, and subsequently detected by the 
SPD with a Gouy phase control using a lens.
Careful balancing of the power and Gouy phase in the two arms ensures arbitrarily good cancellation 
of extraneous noises (acquired by the laser beam before the PBS, \cref{fig:concept}(b)(iii)) and
transduction of torsional motion into transverse displacement at the SPD (\cref{fig:concept}(b)(iv)). 
Thus, for the mirrored optical lever, the photocurrent density,
\begin{equation}\label{PSD2}
    S_{I}[\Omega] = \frac{2}{\pi}(2\eta R P k w_0 \sin{\zeta})^2 \Big[S_{\theta}^{\mathrm{phys}}[\Omega] 
    + \frac{\pi/2}{\eta (\bar{a}k w_0)^2}\csc^2{\zeta}\Big],
\end{equation}
is immune to classical noises and thus quantum-noise-limited in its imprecision (see SI for more details).
The trade-off is double the quantum noise from the mirrored beam. In contrast to interferometric detection of linear motion, our scheme is immune to laser phase noise;
yet, the mirrored optical lever is qualitatively similar to homodyne detection: the quantum-noise-limited imprecision
decreases inversely with optical power ($\bar{a}\propto P$) and the Gouy phase plays the role of the 
homodyne angle. We place the SPD at Gouy phase $\zeta = \pi/2$ which maximizes the angular signal.

We evaluate the suppression of classical spatial mode noise with the mirrored optical lever in both the time and frequency domains.
\cref{fig:concept}(d) illustrates the tilt angle fluctuations of the laser beam, referenced to the flat mirror’s position, recorded over 300 s at a sampling rate of 100 Hz. 
Using the mirrored image produced by the reference arm, the tilt fluctuations (blue curve) are significantly suppressed, reducing the standard deviation from 620 nrad to 73 nrad. 
Further analysis of this suppression was performed in the frequency domain. 
\cref{fig:concept}(e) shows the frequency response of the mirrored optical lever, measured using lock-in amplification with beam tilt modulation introduced via an acousto-optic deflector within a $4f$ lens system (this system, which is part of our calibration method, is detailed in the next section; see Fig. 2(a)). 
The measurements reveal a 60 dB reduction in angular fluctuations across a wide frequency range, up to 100 kHz, when the mirrored arm is employed. 
Residual noise is attributed to factors such as imperfect polarization division, the finite extinction ratio of the PBS ($\sim$3000), and other systematic errors.

\begin{figure}[t!]
    \includegraphics[width= \columnwidth]{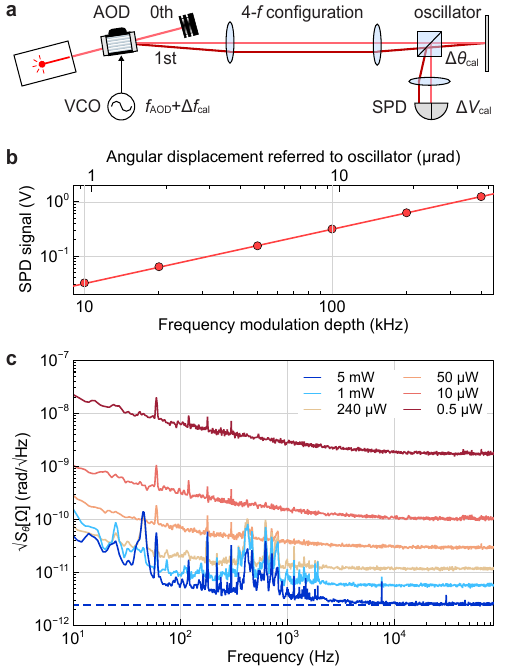}
    \centering
    \caption{\label{fig:calibration}
    (a) Schematic of the acousto-optic deflector (AOD)-based calibration method. The small tilt angle produced by the AOD emulates the angular displacement of the torsional oscillator within the $4f$ lens system.
    (b) Calibration curve for converting the observed SPD signal, measured via the mirrored optical lever (without the reference arm), into tilt angles corresponding to each modulation depth applied to drive the AOD (see text for details).
    (c) Measurement sensitivity of the mirrored optical lever as the optical power is changed. Above
    1 kHz, the observed noise is consistent with quantum vacuum fluctuations (blue dashed line) in the 
    higher-order spatial modes.
    }
\end{figure}

\section{Calibration and performance of mirrored optical lever}
In order to investigate the calibration and sensitivity of the mirrored optical lever, we first operate it with a flat mirror, 
instead of the torsional oscillator, in the signal arm.

Independent calibration of the SPD voltage into angular motion is crucial for further investigation \cite{SunGao21}.
(Note that estimating the angular displacement from the measured optical lever arm does not 
hold beyond the Rayleigh length.)
We perform direct calibration against a known frequency modulation using an acousto-optic deflector (AOD) within a $4f$ lens system. To wit, an AOD is placed at the input plane of the entrance lens of a $4f$ imaging system, with the torsional oscillator (and retroreflector) at the output plane of the second lens (\cref{fig:calibration}(a)).
In this configuration, the tilt of the laser beam at the input plane can be modulated by frequency-modulating the AOD drive as 
$\Delta \theta_{\mathrm{cal}} = (\lambda/v_c) \Delta f_{\mathrm{cal}}$, where $\Delta f_\mathrm{cal}$ is the frequency-modulation
depth, and $v_c \approx \SI{5700}{m/s}$ is the acoustic velocity of the AOD quartz crystal.
This known tilt change manifests as a voltage change ($\Delta V_{\mathrm{cal}}$) in the SPD signal.
By estimating the calibration factor, defined as $\alpha_{\mathrm{cal}}=\Delta \theta_{\mathrm{cal}}/2\Delta V_{\mathrm{cal}}$, the angular displacement spectrum is computed from the SPD spectrum: $S_\theta[\Omega]=\alpha_{\mathrm{cal}}^2 S_V[\Omega]$, where $S_V[\Omega]$ is the measured SPD spectrum (see SI for details).
\cref{fig:calibration}(b) shows the voltage amplitude detected by the SPD with different frequency modulation depths of the AOD drive. 
The beam tilt modulation using the AOD enables a highly linear calibration of the voltage to the angular spectrum, achieving an R-square value of 0.99996 and a calibration factor error of 1.5\%, as determined by the linear fit.

\cref{fig:calibration}(c) demonstrates the calibrated performance of the mirrored optical lever. 
The angle-referred imprecision noise decreases as the reflected optical power increases, and 
reaches $\SI{2.6e-12}{rad/\sqrt{Hz}}$ with 5 mW incident power, consistent with the quantum-noise scaling 
in \cref{PSD2} corresponding to a detection efficiency $\eta \approx 0.75$.

The excess noise peaks between 200 Hz and 1 kHz are attributable to seismic noise-induced 
fluctuations (see SI). 
Without the mirrored arm, we observed extraneous low frequency tilt noise above 1 mW 
of optical power that compromised the linearity of the SPD; with the mirrored arm, this low frequency noise was effectively suppressed, as shown in \cref{fig:concept}(e).
Given that our lab temperature is stabilized to a precision around 20 mK \cite{FifSud24}, we conjecture
that the low frequency drifts in input laser tilt are due to refractive index fluctuations from air currents.

\begin{figure}[t!]
    \includegraphics[width=\columnwidth]{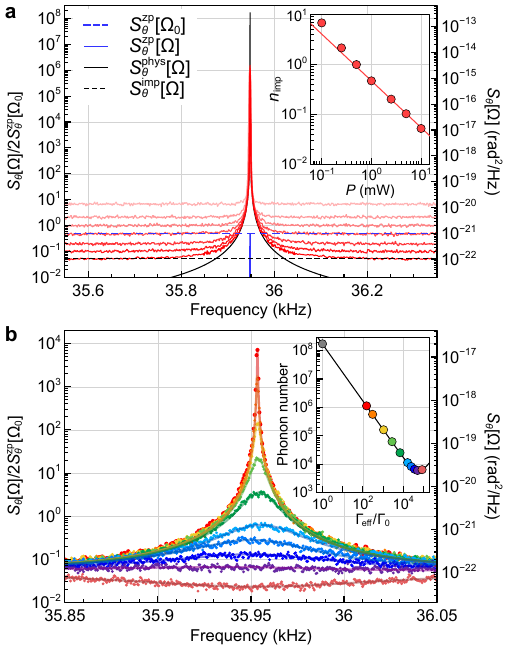}
    \centering
    \caption{\label{fig:motion} 
    (a) Thermal Brownian motion of the fundamental torsional mode measured with increasing optical power. 
    The dashed blue line shows the peak zero-point motion of the fundamental torsional mode. 
    Inset shows the phonon-equivalent imprecision, whose scaling is consistent with quantum-noise-limited detection, 
    and reaches a minimum of $n_\mathrm{imp} \approx 5\times 10^{-2}$.
    (b) At the lowest imprecision, the measurement record is used in a feedback
    loop (see \cref{fig:concept}a) to cool the torsional mode via radiation torque to a final phonon occupation 
    of $n_\mathrm{eff} \approx 6\times 10^3$.}
\end{figure}

\section{Laser cooling of torsional oscillator}
\subsection{Imprecision below the zero-point motion}
Next, we place the torsional oscillator in the signal arm of the mirrored optical lever. 
\Cref{fig:motion}(a) shows the power spectrum of the measured angular fluctuations of the fundamental torsional
mode as the power in the measurement field increases. 
The observed angle fluctuations 
$\delta\hat{\theta}_\mathrm{obs} = \delta\hat{\theta}_\mathrm{phys} + \delta\hat{\theta}_\mathrm{imp}$
consists of the physical motion of the oscillator $\delta\hat{\theta}_\mathrm{phys}$ and the imprecision noise 
$\delta\hat{\theta}_\mathrm{imp}$ of the optical lever. 

The physical motion, shown as the black line in \cref{fig:motion}(a), is the sum of the intrinsic
motion due to thermal and zero-point fluctuations, and a sub-dominant contribution due to the quantum back-action
of the measurement. The intrinsic motion is predominantly the 
thermal motion due to its $n_\mathrm{th} = k_B T/(\hbar\Omega_0) \approx 2\times 10^8$ 
average phonons at room temperature, and is described by the fluctuation-dissipation theorem \cite{CallWelt51}, 
$S_\theta^\mathrm{int}[\Omega] = 4\hbar(n_\mathrm{th} + \tfrac{1}{2}) \text{Im} \chi_0[\Omega]$; here $\chi_0[\Omega] = 
[I(-\Omega^2 + \Omega_0^2 + i \Omega \Gamma_0[\Omega])]^{-1}$ is the susceptibility of the torsional mode
with moment of inertia $I$ and damping rate $\Gamma_0[\Omega] = (\Omega_0/Q)(\Omega_0/\Omega)$.
The quantum back-action of the measurement is primarily due to (see SI) quantum fluctuations in the transverse
displacement of the measurement beam, which cause quantum radiation torque fluctuations on the oscillator,
$S_\tau^\mathrm{ba}[\Omega] \approx 2(\hbar \bar{a}k w_0)^2$. The black line in \cref{fig:motion}(a) is a model
of the physical motion $S_\theta^\mathrm{phys}$ based on independent 
measurement of the frequency, mechanical $Q$, and 
calibrated mode temperature. The latter is inferred by calibrating each spectra using our frequency
modulation technique, and assuming that the torsional mode is in thermal equilibrium.
We verified (see SI) that the mode temperature is constant across all powers except the highest, which shows a
13\% increase, presumably due to optical absorption.

The observed imprecision noise (\cref{fig:motion}(a)) is inversely proportional to the optical power, 
consistent with quantum-noise-limited measurement.
The inset shows the measurement imprecision around resonance in
equivalent phonon units, $n_\mathrm{imp} = S_\theta^\mathrm{imp}/2S_\theta^\mathrm{zp}[\Omega_0]$, where 
$S_\theta^\mathrm{zp}[\Omega_0] = 4\theta_\mathrm{zp}^2/\Gamma_0$ is the peak spectral density of the zero-point motion
$\theta_\mathrm{zp} = \sqrt{\hbar/(2I \Omega_0)}$ of the torsional mode.

The overall low imprecision of the measurement, $n_\mathrm{imp}\approx 5\times 10^{-2}$, is 9.8 dB below
the peak zero-point motion of the fundamental torsional mode. 
This performance can also be compared against the ideal performance achievable in weak continuous measurement
of angular displacements. The minimum total noise achievable in such a measurement, without the use of any
quantum correlations is (see SI):
\begin{equation}
    S_\theta^\mathrm{obs} = S_\theta^\mathrm{int} + S_\theta^\mathrm{ba} + S_\theta^\mathrm{imp}
        \geq S_\theta^\mathrm{zp} + 2\sqrt{S_\theta^\mathrm{ba} S_\theta^\mathrm{imp}}.
\end{equation}
Accounting for the tradeoff
between measurement and its complementary disturbance for the mirrored optical lever scheme (see SI), 
$S_\theta^\mathrm{ba} S_\theta^\mathrm{imp} \geq \sqrt{\pi/2} \cdot 2\hbar \abs{\chi_0}$, gives the standard quantum limit (SQL) 
at every frequency for this measurement scheme. 
On-resonance, the contribution of the zero-point motion and the imprecision-back-action tradeoff is equal, 
giving the particularly simple expression, $S_\theta^\mathrm{SQL}[\Omega_0] \equiv (1+\sqrt{\pi/2})
S_\theta^\mathrm{zp}[\Omega_0] \approx 2.25\cdot S_\theta^\mathrm{zp}[\Omega_0]$. 
That is, \emph{on resonance}, the imprecision noise of our measurement compares favourably against the 
SQL for angular displacement measurement.
(Note that in contrast to prior work \cite{KimDav16}, we achieve this sensitivity at room temperature and in free space; compared to recent work \cite{PraWil23,HaoPur24} 
demonstrating better measurement precision relative to the on-resonance SQL, our work
features better absolute sensitivity due to the mirrored optical lever.)

\subsection{Laser cooling by radiation torque feedback}
The ability to measure with an imprecision far below the peak zero-point motion directly informs the efficacy of
laser cooling of torsional motion, eventually into the ground state, paralleling the development of feedback 
cooling of linear motion \cite{CohPin99,CourPin01,KlecBouw06,PogRug07,WilKip15,RossSch18,WhitSud21,MagAsp21,TebNov21}.
The primary requirement is that the measurement imprecision, relative to the peak zero-point motion, 
be comparable to the inverse of the thermal occupation.

We apply radiation pressure torque ($\delta \tau_\mathrm{fb}$) from a second laser beam, conditioned on the observed motion
$\delta \theta_\mathrm{obs}$, to actuate on the torsional mode.
The observed motion is used to synthesize a signal that drives an amplitude modulator
in the path of an actuation laser. This beam is focused to a $\SI{50}{\mu m}$ spot on the edge of the torsional oscillator 
(at an angle with respect to the measurement beam so as to not cause scatter into the measurement beam, see \cref{fig:concept}(a)).
Torque actuation is exclusively driven by radiation pressure and constrained by imprecision noise at the SPD, ensuring that feedback cooling is governed primarily by the observed motion (see SI).
The resulting optical feedback torque, $\delta \tau_\mathrm{fb} = -\chi_\mathrm{fb}^{-1} \delta \theta_\mathrm{obs}$, can be engineered to
affect damping by adjusting the phase of the feedback filter $\chi_\mathrm{fb}^{-1}$ to be $\pi/2$ around resonance. 
The low imprecision of the measurement guarantees that the damped motion is in fact cold.

The torsional mode is cooled by increasing the gain in the feedback loop. \cref{fig:motion}(b) shows the spectrum $S_\theta^\mathrm{obs}$ of
the observed motion as the gain is increased.
The effective damping rates ($\Gamma_\mathrm{eff}$) are estimated by fitting the curves to a model of the apparent motion (see SI for details
of the model); solid lines show these fits. 
Assuming that the torsional mode satisfies the equipartition principle, we estimate the phonon occupation from the model of the 
physical angular motion, using parameters inferred from fits to the observed motion:
\begin{equation}\label{neff}
    n_{\text{eff}}\approx\int\frac{S_{\theta}^\mathrm{phys}}{2\theta_{\text{zp}}^2}\frac{d\Omega}{2\pi}\approx n_\text{th}\frac{\Gamma_0}{\Gamma_{\text{eff}}}+n_{\text{imp}}\frac{\Gamma_{\text{eff}}}{\Gamma_0}.
\end{equation}
The inset in \cref{fig:motion}(b) shows the inferred phonon occupation as the effective damping rate is increased by the feedback gain.
Feedback cooling cools the torsional mode from an average phonon occupation of $\sim 2\times 10^8$ (at room temperature) to about $5964\pm39$, 
ultimately limited by heating by the feedback of imprecision noise, consistent with \cref{neff}. 
This sets the limit $n_{\text{eff}} \gtrsim 2\sqrt{n_{\text{th}}n_{\text{imp}}}$, which is consistent
with what is observed.

\section{Conclusion}
We demonstrated laser cooling of a centimeter-scale torsion oscillator using a measurement 
whose imprecision is $9.8\, \mathrm{dB}$ below the peak zero-point motion. 
The performance of cooling is limited by the achievable strength of the measurement given the size of the 
zero-point motion. 
This limitation can be overcome in two stages: first, through coherent angular signal amplification \cite{Shimoda22} using a degenerate optical cavity \cite{Arnaud69} where both $\mathrm{HG}_{00}$ and $\mathrm{HG}_{10}$ modes resonate simultaneously, and second, by implementing laser cooling in a strong optical trap \cite{WhitSud21}, where the requirements are significantly reduced \cite{KomSud22} compared to the current protocol.
Thus, our work opens the door for eventual quantum state control of macroscopic torsional pendula. 
Combined with the pedigree and advantage of torsional pendula in the measurement of weak gravitational 
interactions, this work establishes the necessary step toward mechanical experiments that explore the 
fundamental nature of gravity. 

\let\oldaddcontentsline\addcontentsline
\renewcommand{\addcontentsline}[3]{}

\bibliography{../../refs_torsion_pendulum}

\let\addcontentsline\oldaddcontentsline


\clearpage
\onecolumngrid
\appendix 

\begin{center}
    \large{\bfseries Supplementary Information}
\end{center}

\tableofcontents

\section{Torsion pendulum}

\subsection{Fabrication}

\begin{figure}[h!]
    \centering
    \includegraphics[width=0.9\columnwidth]{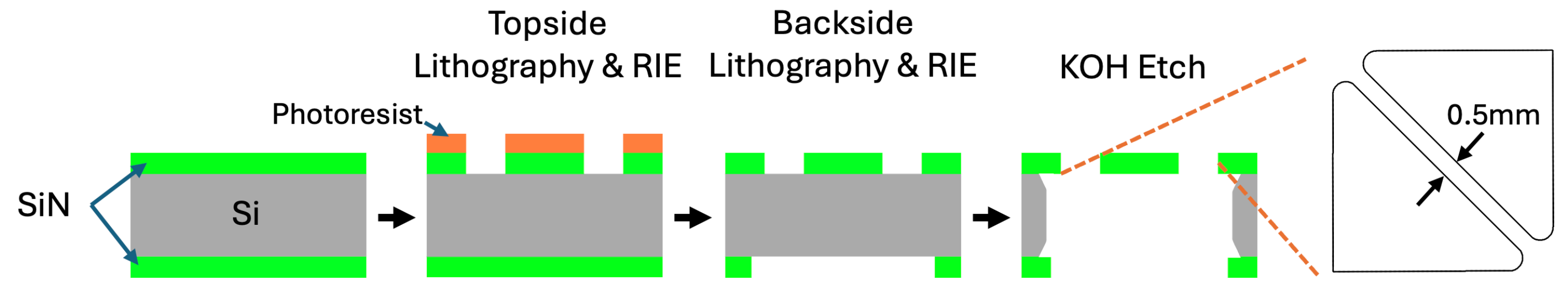}
    \caption{\label{fig:fab}Essential steps in the process flow for fabricating the torsion pendulum.}
\end{figure}

The torsion pendulum used in this work is fabricated starting
with a double-sided, 400 nm thick $\t{Si_3 N_4}$-on-Si wafer (WaferPro); 
\cref{fig:fab} shows the essential steps in the process.
We coated both sides with $\SI{3}{\mu m}$ thick S1813 positive photoresist. Using mask lithography, 
we patterned the top-side resist with the pendulum shape, while the bottom-side resist protected the wafer from 
scratches during handling.
We transferred the pattern to the SiN film using a $\t{CF_4}-$based reactive ion etch. After removing the 
remaining photoresist, we applied a fresh coat to both sides. We then patterned the backside with a square window, 
carefully aligning it with the front pattern. After dry-etching the square window into the wafer, we applied a thick layer of photoresist to both sides for protection. We diced the wafer into individual samples. We removed the photoresist using 
an acetone, iso-propyl alcohol (IPA), and de-ionized (DI) water rinse, followed by an oxygen plasma clean to eliminate any 
lingering residue. We vertically placed each sample in a $74^\circ\, \t{C}$ potassium hydroxide bath for 24 hours. 
After etching, we rinsed the samples in DI water, followed by IPA, and let them air-dry. 
Finally, we gave the device a final oxygen plasma clean to remove any remaining residue.

\subsection{Ringdown measurements}

\begin{figure}[h!]
    \centering
    \includegraphics[width=0.4\columnwidth]{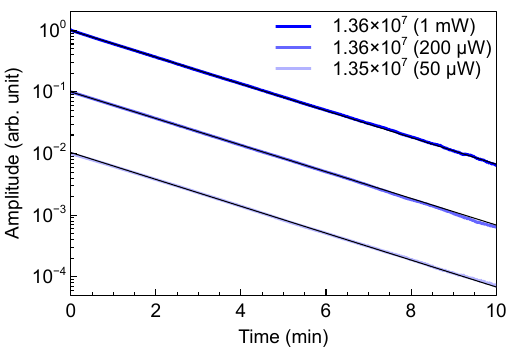}
    \caption{\label{fig:ringdown}Ringdown measurements of the fundamental torsion mode at $\Omega_0/2\pi = 35.95 \t{kHz}$ with different probe beam powers}
\end{figure}

The mechanical Q of the fundamental torsional mode was estimated from ringdown measurements. The motion is
excited by a radiation pressure torque, and measured using the SPD. The measured signal is demodulated at the
mechanical frequency using a lock-in amplifier (Moku:Pro, Liquid Instruments).
\Cref{fig:ringdown} shows such measurements at a few different powers of the measurement beam.

\section{Noise in an optical lever}

To better understand the quantum noises in optical lever detection, we consider here the decomposition of a 
quantized optical field on an orthonormal Hermite-Gaussian (HG) basis. 
The HG mode $U_{mn}$ of \(m\)-th order in x-axis and \(n\)-th order in y-axis is defined as \cite[\S 16.4]{Sieg86}
\begin{align}
    U_{mn}(x,y,z,t) &= u_{mn}(x,y,z)e^{i\phi_{mn}(x,y,z,t)},\\
    u_{mn}(x,y,z) &= \sqrt{2\over{\pi}}{1\over{w(z)}}{1\over{\sqrt{2^{m+n}m!n!}}}e^{-(x^2+y^2)/w^2(z)} 
        H_{m}(\sqrt{2}x/w(z)) H_n\sqrt{2}y/w(z)),\\
    \phi_{mn}(x,y,z,t) &= \omega_\ell t-kz-\frac{k(x^2+y^2)}{2R(z)}+(m+n+1)\zeta(z);
\end{align}
here, \(\omega_\ell\) and \(k\) are the angular frequency and wavenumber of the laser beam; 
\(w(z)=w_{0}\sqrt{1+(z/z_{R})^2}\), \(R(z)=z(1+(z_{R}/z)^2)\), and \(\zeta(z)=\text{artan}(z/z_{R})\) are the beam radius, radius of curvature, and Gouy phase, respectively; 
\(w_{0}\) is the beam width at the waist; \(z_{R}=kw_{0}^2/2\) is the Rayleigh length; and
$H_n$ is the $n^\t{th}$ Hermite polynomial. 
These bases satisfy the orthonormality relation
\(\int \dd x \dd y\, U_{mn}^* U_{m'n'}=\delta_{mm'}\delta_{nn'}\).

The quantized electric field (in an arbitrary, but static polarization direction) of a laser beam with
power $P$ in the (fundamental) $\t{HG}_{00}$ mode --- such as the emission of an ideal laser --- can be modeled by
\begin{equation}
    \hat{E}(x,y,z,t) = (\bar{a}+ \dop{a}_{00}(t)) U_{00}+\sum_{mn}^{}\delta\hat{a}_{mn}(t)U_{mn}.
\end{equation}
Here \(\bar{a}=\sqrt{P/\hbar \omega_\ell}\) is the mean photon flux amplitude and the operators 
\(\delta \hat{a}_{mn}\) represent fluctuations in \(U_{mn}\) mode, that satisfy the 
canonical commutation relations \cite{Wun04}:
\begin{equation}
    \left[\delta\hat{a}_{mn}[\Omega], \delta\hat{a}_{m'n'}[\Omega']^{\dagger}\right]
    = 2\pi \delta_{mm'}\delta_{nn'}\delta[\Omega+\Omega'].\label{comm}
\end{equation}
In the ideal case, the fluctuations are given by vacuum fluctuations of the HG modes, $\dop{a}_{mn}^{0}$, derived through the projection of the fluctuations of a plane wave into the HG modes. 
To account for extraneous transverse displacement and tilt noises, we set 
\begin{equation}
    \dop{a}_{mn} = \dop{a}_{mn}^{0}  + \delta a_{mn}^{\t{ext}},
\end{equation}
where $\delta a_{mn}^\t{ext}$ represent the extraneous fluctuations in each mode.

Let us consider a case where the laser beam at arbitrary point \(z=Z\) is transversely displaced by \(\delta x\) and tilted by \(\delta \theta\) in the xz plane, viz., 
\begin{equation}
    \hat{E}(x,y,Z,t) \rightarrow \hat{E}(x-\delta x,y,Z,t)e^{ik\delta\theta x} 
    \approx \hat{E}(x,y,Z,t)+\bar{a}\left(\frac{1}{w(Z)}\delta \hat{x}+i\frac{kw(Z)}{2}\delta \hat{\theta} \right)
    e^{-i\zeta(Z)}U_{10}, \label{scattering}
\end{equation}
where we assume \(\delta x, \delta \theta \ll 1\), and have then represented these fluctuations by associating
them with operators.
\cref{scattering} shows that the laser beam's transverse displacement and tilt can be interpreted as the scattering of the fundamental mode to the first-order mode. For instance, at the beam waist (\(z=0\)), the displacement (tilt) fluctuations are encoded into the amplitude (phase) quadrature of the first-order mode, while in the far field (\(z\ll z_0\) or \(z\ll-z_0\)) the fluctuations are transposed to the opposite quadratures. Meanwhile, \cref{scattering} also explicitly shows how the first-order amplitude and phase quadratures contribute to the displacement and tilt fluctuations along the beam propagation:
\begin{align}
    &\dop{x}(Z) = \frac{w(Z)}{2\bar{a}}(\delta\hat{a}_{10} e^{i\zeta(Z)}+\delta\hat{a}_{10}^\dagger e^{-i\zeta(Z)})\label{displacement}\\ 
    &\dop{\theta}(Z) = \frac{1}{i\bar{a}kw(Z)}(\delta\hat{a}_{10} e^{i\zeta(Z)}-\delta\hat{a}_{10}^\dagger e^{-i\zeta(Z)}) \label{theta}
\end{align}
This represents the essence of the optical lever technique; the laser tilt produced by a torsion pendulum at the beam waist is converted into the transverse displacement at the far field.

\subsection{Measurement imprecision in split photo-detected optical lever}

Thus, a split photodetector (SPD) can used in the far field to detect the transverse displacement of the laser beam,
and thereby infer the tilt that produced it \cite{EnoKawa16,HaoPur24}. 
Specifically, the SPD takes the difference of the photocurrents from a pair of photodetectors placed next to each
other, which we take to be horizontal in the plane perpendicular to the propagation direction $z$. 
The photocurrent operator \(\hat{I}(t;z)\) for the SPD located at a distance \(z\) can be computed by the photon number operator, \(\hat{n}=\hat{a}^\dagger \hat{a}\), as
\begin{equation}
    \begin{split}
        \hat{I}(t;z)& = q_e \left[\hat{n}_{\text{R}}(t)-\hat{n}_{\text{L}}(t)\right]\\
        &=q_e\int_{-\infty}^{\infty}dx\int_{-\infty}^{\infty}dy \ \text{sgn}(x)\hat{E}^{\dagger}(x,y,z,t)\hat{E}(x,y,   z,t)\\
        &=2q_e\bar{a}\sum_{m,n\neq0}^{}\frac{1}{\sqrt{2^{m+n}m!n!}}\Re\left[\delta\hat{a}_{mn}(t)e^{i(m+n)\zeta(z)}\right]\int_{-\infty}^{\infty}dx \ \text{sgn}(x) \int_{-\infty}^{\infty}dy \frac{2}{\pi w(z)^2} H_m(x)H_n(y)e^{-\frac{2(x^2+y^2)}{w(z)^2}}\\
        &=\frac{2}{\pi}q_e\bar{a}\sum_{m,n\neq0}^{}\frac{1}{\sqrt{2^{m+n}m!n!}}\Re\left[\delta\hat{a}_{mn}(t)e^{i(m+n)\zeta(z)}\right]\int_{-\infty}^{\infty}dx \ \text{sgn}(x) H_m(x)e^{-x^2}\int_{-\infty}^{\infty}dy H_n(y)e^{-y^2}\\
        &=2q_e\bar{a}\sum_{m\neq0}^{}\frac{1}{\sqrt{\pi2^{m}m!}}\Re[\delta\hat{a}_{m,0}(t)e^{im\zeta(z)}]\int_{0}^{\infty}dx \ (1-(-1)^m) H_m(x)e^{-x^2}\\
        &=4q_e\bar{a}\sum_{k=0}^{\infty}\frac{1}{\sqrt{\pi2^{2k+1}(2k+1)!}}\Re[\delta\hat{a}_{2k+1,0}(t)e^{i(2k+1)\zeta(z)}]\int_{0}^{\infty}dx H_{2k+1}(x)e^{-x^2}\\
        &=4q_e\bar{a}\sum_{k=0}^{\infty}\Re[\delta\hat{a}_{2k+1,0}(t)e^{i(2k+1)\zeta(z)}]\frac{(-1)^k}{(2k+1)k!}\sqrt{\frac{(2k+1)!}{\pi 2^{2k+1}}}\\
        &=\frac{2}{\sqrt{\pi}}q_e\bar{a}\sum_{k=0}^{\infty}\left[\delta\hat{q}_{2k+1,0}(t)\cos{((2k+1)\zeta(z))}-\delta\hat{p}_{2k+1,0}(t)\sin{((2k+1)\zeta(z))}\right]D_k\\ \label{Ispd}
    \end{split}
\end{equation}
where \(q_e\) is the electron charge, \(\delta\hat{q}_{mn}(t)=(\delta\hat{a}_{mn}+\delta\hat{a}_{mn}^{\dagger})/\sqrt{2}\) and \(\delta\hat{p}_{mn}(t)=(\delta\hat{a}_{mn}-\delta\hat{a}_{mn}^{\dagger})/\sqrt{2}i\) are amplitude and phase quadrature operators, and the coefficient \(D_k\) is defined as \(D_k =\frac{(-1)^k}{(2k+1)k!}\sqrt{\frac{(2k+1)!}{2^{2k}}}\). \cref{Ispd} shows that the split photodetection leads to the detection of quadratures of the odd-number HG modes \(\delta\hat{a}_{2k+1,0}\), being amplified by the amplitude $\bar{a}$ of the fundamental HG mode, and with a rotation determined by the Gouy phase acquired during propagation. 
That is, in analogy with a balanced homodyne detection employed for measurement of longitudinal phase
fluctuations, the amplitude of fundamental mode \(\bar{a}\) and the Gouy phase shift act as the local 
oscillator beam and the homodyne angle, respectively.

Suppose now that the angular fluctuations are produced by the motion of a torsion pendulum positioned at \(z=z_0\). 
The laser beam reflected off from the torsion pendulum, whose angular motion is denoted as $\dop{\theta}_{\t{sig}}$, attains a tilt of $2\dop{\theta}_{\t{sig}}$. Given that the laser beam is subject to extraneous classical noises in the first-order HG mode, i.e. classical transverse displacement ($\dop{x}_{\t{ext}}$) and tilt noise ($\dop{\theta}_{\t{ext}}$) referred to the position of the torsion pendulum, the optical field can be expressed as
\begin{equation}
    \hat{E}(\vb{r},t) \approx \bar{a}U_{00}(\vb{r},t)+\Big[ \frac{\bar{a}}{w(z_0)}\dop{x}_{\t{ext}}+i\frac{kw(z_0)}{2}(2\dop{\theta}_{\t{sig}}+\dop{\theta}_{\t{ext}})\Big]U_{10}(\vb{r},t)+ \sum_{n,m} \dop{a}_{nm}(t) U_{nm}(\vb{r},t). \label{field}
\end{equation}
Then, the symmetrized single-sided spectral density of the photocurrent can be computed from \cref{Ispd} as
\begin{equation}
    S_{I}[\Omega] = \frac{2}{\pi}(2\eta R P k w(z_0) \sin{\zeta})^2\Big[S_{\theta}^{\t{sig}}[\Omega] + \frac{1}{4}S_{\theta}^{\t{ext}}[\Omega] + \left( \frac{\cot{\zeta}}{k w(z_0)^2} \right)^2 S_{x}^{\t{ext}}[\Omega]+ \frac{\pi/2}{2\eta (\bar{a}k w(z_0))^2}\csc^2{\zeta}\Big],\label{Ipsd}
\end{equation}
where \(R= q_e/\hbar w_l\) and \(\eta\) are the responsivity and the quantum efficiency of the SPD, and \(\zeta=\zeta(z)-\zeta(z_0)\) is the Gouy phase shift from the torsion pendulum to the SPD. In this calculation, the 
single-sided spectral densities of the vacuum fluctuations are \(S_{q}^{mn,0}=1\) and \(S_{p}^{mn,0}=1\) 
(\cref{comm}), and \(\sum_{k=0}^{\infty}D_k^2=\pi/2\). 
Note that transduction of angular motion to SPD signal reaches its maximum value when the Gouy phase shift is set to 
\(\pi/2\). This condition can be achieved either by preparing a sufficient lever length, i.e., \(z-z_0 \gg z_R\), or by controlling the accumulated Gouy phase shift with a lens.
The last term in \cref{Ipsd} gives the imprecision noise due to quantum fluctuations in the higher-order modes,
\begin{equation}\label{theta_imp}
    S_{\theta}^{\text{imp}}[\Omega] 
    = \frac{\pi/2}{2\eta (\bar{a}k w(z_0))^2}\csc^2{\zeta},
\end{equation}
whereas the second and third terms represent the imprecision noise from extraneous fluctuations in the laser.

\subsection{Measurement back action in optical lever}
We consider the quantum back-action torque induced by the laser beam in the optical lever. In the main manuscript, it is assumed that the laser beam is incident at the center of the torsion pendulum. Here, we address the scenario where the laser beam is offset by a distance \(x_0\) from the center of rotation of the torsion pendulum. By linearizing the torque fluctuations, the back-action torque can be expressed as
\begin{equation}
    \delta \hat{\tau}_{\text{ba}}[\Omega] = F{\delta}\hat{x}[\Omega] + \delta \hat{F}[\Omega] x_0,
\end{equation}
where the first term represents the back-action torque due to the transverse displacement fluctuation, and the second term accounts for the force fluctuation. The spectral density of the back-action torque is given by
\begin{equation}
    S_{\tau}^{\text{ba}}[\Omega] = F^2S_{x}[\Omega] + x_0^2 S_{F}[\Omega].
\end{equation}
In general, these fluctuations can be decomposed into two contributions: extraneous classical noise and quantum noise. Therefore, we incorporate the classical components of the fluctuations, denoted as \(S_{x}^{\t{ext}}[\Omega]\) and \(S_{F}^{\t{ext}}[\Omega]\). The quantum fluctuation of the transverse displacement can be derived from Eq.\eqref{displacement} as \(S_{x}[\Omega] = w^2(z_0)/2\bar{a}^2\), while the quantum radiation pressure force fluctuation is given by \(S_{F}^{\text{ba}}[\Omega] = 8\bar{a}^2\hbar^2k^2\), with the mean radiation pressure force \(F_{\t{ba}} = 2\bar{a}^2\hbar^2k^2\). Consequently, the total back-action torque is expressed as
\begin{equation}\label{Sba}
    S_{\tau}^{\text{ba}}[\Omega] = (\sqrt{2}\hbar\bar{a}kw(z_0))^2\left[1 + \left(\frac{2x_0}{w(z_0)}\right)^2\right] + (2\hbar\bar{a}^2k)^2S_{x}^{\t{ext}} + x_0^2 S_F^{\t{ext}}.
\end{equation}
The first term arises from quantum fluctuations in the optical beam used for the measurement (with the term $\propto x_0$ coming from any
potential mis-centering of the beam on the torsion pendulum), while the second and third terms arise from extraneous fluctuations in it.

\subsection{Total measurement noise and the Standard Quantum Limit}

The observed SPD signal, referred to angle, can be understood as the sum of the intrinsic 
(i.e. thermal and zero-point) motion, back-action-driven motion, and measurement imprecision, i.e.
\begin{equation}\label{thetaObs}
    \dop{\theta}_\t{obs} = \dop{\theta}_\t{int} + \dop{\theta}_\t{ba} + \dop{\theta}_\t{imp}.
\end{equation}
The intrinsic motion is given by the 
fluctuation-dissipation theorem \cite{CallWelt51}
\begin{equation}
    S_\theta^\t{int}[\Omega] = 4\hbar(n_\t{th}[\Omega]+\tfrac{1}{2}) \Im \chi_0 [\Omega],
\end{equation}
where $n_\t{th}[\Omega] = (e^{\hbar \Omega/k_B T}-1)^{-1}$ is the average thermal phonon occupation at temperature
$T$, and $\chi_0[\Omega] = [I (-\Omega^2 + \Omega_0^2 - i \Omega \Gamma_0[\Omega])]^{-1}$ is the mechanical 
susceptibility for the torsion pendulum with moment of inertia $I$ and 
frequency-dependent structural damping rate \(\Gamma_{0}[\Omega] = (\Omega_0/Q)(\Omega_0/\Omega)\).
The back-action-driven motion is $\dop{\theta}_\t{ba} = \chi_0 \delta \tau_\t{ba}$, whose spectrum is
\begin{equation}
    S_\theta^\t{ba} = \abs{\chi_0}^2 S_\tau^\t{ba},
\end{equation}
with $S_\tau^\t{ba}$ given by \cref{Sba}.

Note from \cref{theta_imp,Sba} that as the measurement imprecision due to quantum noise decreases ($\propto 1/P$), 
quantum back-action increases ($\propto P$). 
Thus the measured angular noise has a minimum
\begin{equation}
\begin{split}
    S_\theta^\t{obs} 
    = S_\theta^\t{int} + S_\theta^\t{ba} + S_\theta^\t{imp}
    \geq S_\theta^\t{int} + 2\sqrt{S_\theta^\t{ba} S_\theta^\t{imp}},
\end{split}
\end{equation}
whose ideal limit (zero temperature, perfect detection efficiency, beam perfectly centered, no extraneous 
imprecision or back-action) on resonance defines the (resonant) standard quantum limit (SQL) of the measurement: 
\begin{equation}
    S_\theta^\t{SQL}[\Omega_0] = 2 \hbar \Im \chi_0[\Omega_0] (1+\sqrt{\pi})
    = 2 S_\theta^\t{zp}[\Omega_0] \frac{1+\sqrt{\pi}}{2}.
\end{equation}
Here the spectral density of the zero-point motion of the torsional mode is
\begin{equation}
    S_\theta^\t{zp}[\Omega_0] = S_\theta^\t{int}[\Omega_0]\vert_{T=0} = \frac{4\theta_\t{zp}^2}{\Gamma_0},
\end{equation}
where $\theta_\t{zp} = \sqrt{\hbar/(2I \Omega_0)}$ is the zero-point motion of the mode.
In contrast to the SQL for the measurement of linear motion, this is larger than twice the resonant 
zero-point motion of the mechanical oscillator by $(1+\sqrt{\pi})/2 \approx 1.386.$

\section{Extraneous noise cancellation in mirrored optical lever}

\begin{figure}
    \centering
    \includegraphics[width=0.9\columnwidth]{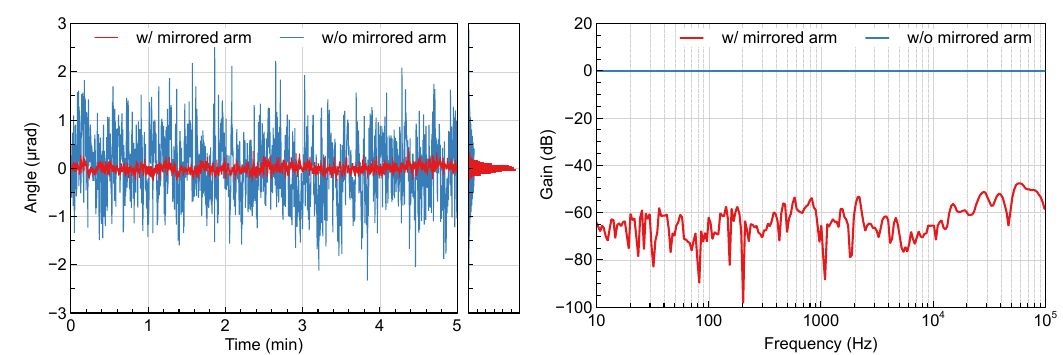}
    \caption{\label{fig:mirrored}
    Characterization of the mirrored optical lever. (a) Ambient angle fluctuation over time with and without the mirrored arm. (b) Frequency response measurement. A sinusoidal tilt modulation is produced by the acousto-optic deflector and measured by the split photodetector, followed by a lock-in amplification to measure the gain between the input and output signals. The response gain is normalized by the gain without the mirrored arm.
    }    
\end{figure}
Here we describe the noise spectral density in the mirrored optical lever. In this configuration (as shown in Fig. 1 in the main text), the input laser beam is prepared to have equal components for polarizations in S and P, so that the P-polarized component passes on to the torsion pendulum to sense the angular motion (signal arm), while the S-polarized component is diverted to a retroreflector (reference arm). A retroreflector is a \(90^\circ\) corner cube that inverts the tilt and transverse displacement of the beam with respect to the normal reflection from a flat mirror, i.e., it alters the sign of the amplitude flux of the first-order HG mode as \(\delta\hat{a}_{10}(t) \rightarrow -\delta\hat{a}_{10}(t)\). In each arm, a quarter-wave plate is positioned so that both beams are collected from the output port of the PBS.
The output optical field is written as \(\vb*{\hat{E}}_{\t{out}}(\vb{r},t) = \hat{E}_{\text{sig}}(\vb{r},t)\vb*{e}_S+\hat{E}_{\text{ref}}(\vb{r},t)\vb*{e}_P\), where \(\vb*{e}_S\) and \(\vb*{e}_P\) are the unit vectors for S and P polarizations, respectively. The signal and reference fields are obtained from the Eq.\eqref{scattering} as
\begin{gather}
    \hat{E}_{\t{sig}}(\vb{r},t) = \bar{a}U_{00}^{\t{sig}}(\vb{r},t)+\Big[ \frac{\bar{a}}{w(z_0)}\dop{x}_{\t{ext}}+i\frac{kw(z_0)}{2}(2\dop{\theta}_{\t{sig}}+\dop{\theta}_{\t{ext}})\Big]U_{10}^{\t{sig}}(\vb{r},t)+ \sum_{n,m} \dop{a}_{nm}(t) U_{nm}^{\t{sig}}(\vb{r},t),\\
    \hat{E}_{\t{ref}}(\vb{r},t) = \bar{a}U_{00}^{\t{ref}}(\vb{r},t)+\Big[ -\frac{\bar{a}}{w(z_0)}\dop{x}_{\t{ext}}-i\frac{kw(z_0)}{2}\dop{\theta}_{\t{ext}}\Big]U_{10}^{\t{ref}}(\vb{r},t)+ \sum_{n,m} \dop{a}_{nm}(t) U_{nm}^{\t{ref}}(\vb{r},t).
\end{gather}
Here, \(U^{\text{sig(ref)}}_{mn}\) represents the notation for the HG modes coming from the signal (reference) arm.
We assume that the optical power and arm length for each arm are identical.
The combined field is then detected by an SPD, yielding the photocurrent density as
\begin{equation}\label{theta_imp}
    S_{I}[\Omega] = \frac{2}{\pi}(2\eta R P k w(z_0) \sin{\zeta})^2 \Big[S_{\theta}^{\t{sig}}[\Omega] 
    + \frac{\pi/2}{\eta (\bar{a}k w_0)^2}\csc^2{\zeta}\Big],
\end{equation}
This indicates that the reference beam cancels out the classical displacement and tilt noises with the trade-off of the doubled shot noise. The orthogonal polarizations for the optical fields allow us to obviate the requirement for stabilizing laser frequency noise. The sensitivity for the angular motion detection is maximized when the Gouy phase shift \(\tilde{\zeta}\) is precisely set to \(\pi/2\) as follows:
\begin{equation}\label{PSD3}
    S_{I}[\Omega] = \frac{2}{\pi}(2\eta R P k w(z_0))^2 \Big[S_{\theta}^{\t{sig}}[\Omega] 
    + \frac{\pi/2}{\eta (\bar{a}k w(z_0))^2}\Big],
\end{equation}
This result shows that, in principle, the mirrored optical lever with the use of a retroreflector is immune to the classical noises that the laser beam experiences before the PBS.

In our experiment, the output laser beam is split by a knife-edge right-angle mirror and then collected by a low noise balanced photodetector (HBPR-100M-60K-IN-FS, Femto). The difference in photocurrents between the two photodiodes of the BPD corresponds to the transverse displacement of the laser beam at the knife-edge mirror. The Gouy phase shift in Eq. \eqref{theta_imp} is defined by the locations of the torsion pendulum and knife-edge mirror.  We set the Gouy phase to be \(\pi/2\) by manipulating a lens between the torsion pendulum and the knife-edge mirror.

To characterize the classical noise suppression of the mirrored optical lever, we measure the angular fluctuations over time with and without the mirrored arm. In this proof-of-concept experiment, the torsion pendulum is replaced with a flat mirror so that the performance is only limited by the measurement system. \cref{fig:mirrored}(a) shows the angle fluctuations of the laser beam referred to the location of the flat mirror over 300 s with a sampling rate of 100 Hz. We observe that the tilt fluctuation (blue curve) is suppressed when the mirrored image is simultaneously incident on the SPD, reducing the standard deviation from 620 nrad to 73 nrad.
Furthermore, we characterize the mirrored optical lever system in terms of the frequency response. To this end, we generate the periodic virtual tilt signal swept over 1 Hz – 100 kHz using the acousto-optic deflector. This virtual angular signal is detected by the SPD and subsequently lock-in amplified. As shown in \cref{fig:mirrored}(b), we confirm that the mirrored optical lever is capable of reducing the angle fluctuations by 50–60 dB in a broad frequency range up to 100 kHz. Thine residual noise may be due to the imperfect polarization division, the finite extinction ratio of the PBS \(\sim\) 3000, and other systematic errors.

\begin{figure}
    \centering
    \includegraphics[width=0.5\columnwidth]{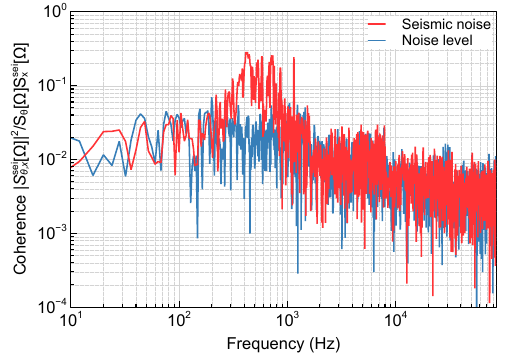}
    \caption{\label{fig:seismic}
    Coherence spectrum between angular displacement and seismic noise.
    }    
\end{figure}

In Fig. 2(a) of the main text, excess noise is observed in the low-frequency regime below 1 kHz Fourier frequency. Potential sources of this extraneous noise include seismic noise and residual spatial noise caused by air current-induced refractive index fluctuations. (Laser intensity noise is significantly suppressed by balanced photodetection at the SPD, and the optical lever is insensitive to the phase quadrature of the laser beam.) To elucidate the origins of this excess noise, we examine the coherence spectrum between the angular displacement noise $S_{\theta}[\Omega]$ and seismic displacement noise $S_{x}^{\t{sei}}[\Omega]$. For this purpose, seismic acceleration $S_{a}^{\t{sei}}$ is measured and low-pass filtered to obtain $S_{x}^{\t{sei}}[\Omega]=S_{a}^{\t{sei}}[\Omega]/\Omega^4$.

\Cref{fig:seismic} shows the coherence spectrum, $|S_{\theta x}^{\t{sei}}[\Omega]|^2/S_{\theta}[\Omega]S_{x}^{\t{sei}}[\Omega]$, between angular displacement and seismic noise. A relatively high coherence is observed in the range of 300 Hz – 1 kHz, resembling the excess noise seen in Fig. 2(a) of the main text. Therefore, the noise peaks in this frequency range are likely attributable to seismic noise, while the long-term noise below 300 Hz may be caused by air current-induced refractive index fluctuations.

\section{Angle calibration}

\subsection{Calibration of mirrored optical lever}
\begin{figure}
    \centering
    \includegraphics[width=\columnwidth]{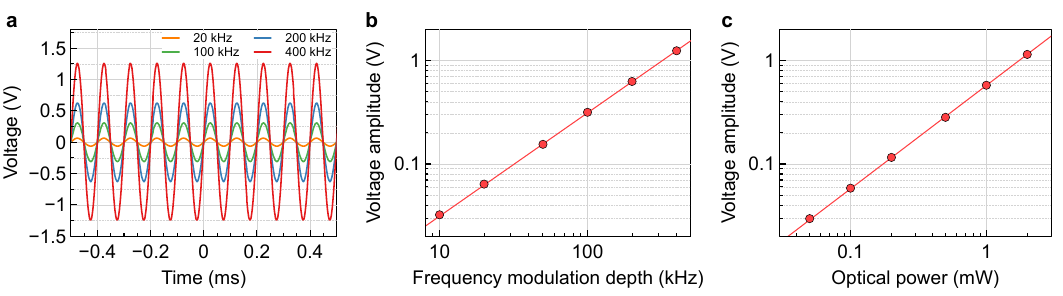}
    \caption{\label{fig:AOD}
    Characterization of the acousto-optic deflector (AOD)-based voltage-to-angle calibration.
    (a) Sinusoidal frequency modulation of the AOD drive leads to a corresponding modulation in the signal detected by the single-photon detector (SPD) in the time domain. To minimize extraneous noise and improve the accuracy of the voltage amplitude estimation, the SPD signal is triggered and averaged over 100 iterations.
    (b) The relationship between the frequency modulation depth (corresponding to beam tilt amplitude) and the SPD signal is linear, with a coefficient of determination $R^2= 0.9996$. The incident optical power on the SPD is $\SI{500}{\micro W}$.
    (c) A linear relationship also exists between the optical power and the SPD signal amplitude, with the AOD drive modulation depth set to 100 kHz.
    }
\end{figure}

In conventional optical lever techniques, the angular displacement of the torsion pendulum, $\delta\theta$, is inferred from the geometry of the lever: a laser beam deflected with a tilt of $2\delta\theta$ propagates over a distance $z$, thereby amplifying the transverse displacement by the lever arm, i.e., $\delta x = 2z\delta\theta$. 
This suggests that extending the lever arm could infinitely amplify the SPD signal gain. 
However, this naive calibration is only valid within the framework of ray optics; beyond the Rayleigh length of the laser beam, the gain of the SPD signal eventually saturates.

For accurate measurement of the intrinsic thermal motion of a high-$Q$ torsion pendulum, calibration is often performed by fitting the theoretical model of thermal motion to the SPD voltage spectrum. 
However, this method can lead to inaccuracies when extraneous noise, such as photothermal effects or external disturbances, contaminates the angular displacement spectrum.

To overcome this limitation, we employ a direct calibration method that relates the SPD signal to angular displacement using an acousto-optic deflector (AOD). 
As described in the main text, an AOD (M1377-aQ80L-1, ISOMET) is positioned at the input plane of the entrance lens of a $4f$ imaging system, with the torsion pendulum (and retroreflector) located at the output plane of the second lens. 
The $4f$ system, composed of two lenses with identical focal lengths, reproduces the angular tilt of the laser beam at the input plane at the output plane.

We use the first-order diffraction beam from the AOD, with its tilt angle precisely controlled by the relation $\Delta \theta_{\t{cal}} = (\lambda/v_c) \Delta f_{\t{cal}}$, where $\Delta f_\t{cal}$ is the drive frequency change and $v_c \approx \SI{5700}{m/s}$ is the acoustic velocity of the AOD quartz crystal.
Using this relation, we determine the calibration factor for converting the SPD voltage spectrum into an angular displacement spectrum.
Since the laser beam acquires an angular displacement that amounts to twice the torsional motion of the pendulum upon reflection, the calibration factor is defined as $\alpha_{\t{cal}}=\Delta \theta_{\t{cal}}/2\Delta V_{\t{cal}}$, where $\Delta V_{\t{cal}}$ represents the voltage change in the SPD signal.
To precisely attain the calibration factor, we frequency modulate the AOD drive with a known depth ($\Delta f_{\t{cal}}$), and observe the corresponding SPD voltage modulation amplitude ($\Delta V_{\t{cal}}$) in the time domain. 
In this process, we trigger and average the calibration tone 100 times to mitigate extraneous noises in the SPD signal. 
Then, the calibration factor is used to convert the SPD spectrum to the angular displacement spectrum: $S_\theta[\Omega]=\alpha_{\t{cal}}^2 S_V[\Omega]$, where $S_V[\Omega]$ is the measured SPD spectrum. 
Our typical choice for the AOD calibration is a 10 kHz frequency modulation tone with a 100 kHz frequency depth (the corresponding tilt amplitude of $\SI{18.7}{\micro rad}$), centered around 80 MHz (AOD operation frequency).
Note that, although this calibration tone can be detected in the power spectral density (PSD), the measured voltage amplitude may be inaccurate or underestimated due to the broadening of the linewidth in the PSD analysis.

\cref{fig:AOD} illustrates the application of AOD-based calibration in our experiments. 
We vary the frequency modulation depth of the AOD drive and monitor the SPD signal in the time domain. The triggering and averaging of the modulation signal is performed by an oscilloscope (Moku:Pro, Liquid Instruments).
The voltage amplitude is then estimated from these averaged data. 
\cref{fig:AOD}(b) displays the voltage amplitudes as a function of modulation depths, corresponding to beam tilt amplitudes, which provides the calibration factor for converting the SPD spectrum into the angular displacement spectrum. 
The linear relationship is confirmed with a coefficient of determination $R^2 = 0.99996$. 
Additionally, \cref{fig:AOD}(c) demonstrates the linear dependence of voltage amplitude on incident optical power.


\begin{figure}
    \centering
    \includegraphics[width=0.9\columnwidth]{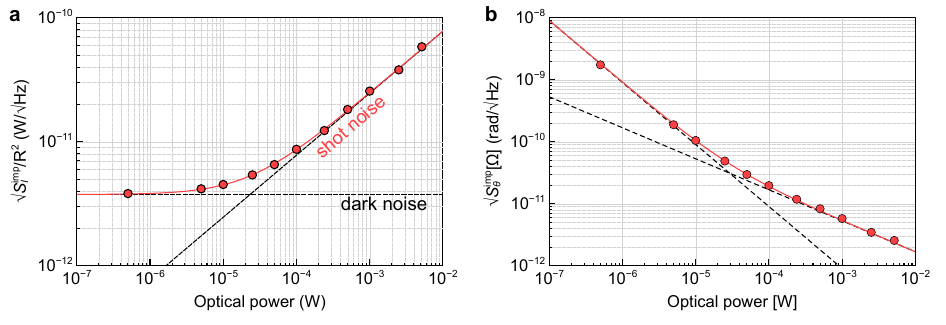}
    \caption{\label{fig:photocurrent}
    (a) Input-referred photocurrent noise. (b) imprecision noise in angular displacement.
    }
\end{figure}
Figure 2(b) in the main text is calibrated from the SPD voltage spectrum to the angular displacement spectrum using the calibration method described above. 
To verify that the imprecision noise in the spectrum is quantum-noise-limited, we measure the noise floor of the photocurrent density, referred back to the optical power, at a frequency of $\Omega = 2\pi\cdot\SI{80}{kHz}$ under varying optical power levels.
As shown in \cref{fig:photocurrent}(a), at power levels below 20 µW, the dark noise is dominant, while shot noise starts to prevail beyond this threshold. 
The curve fitting reveals that the quantum efficiency of the balanced photodetector used in this study is 0.81. 
\cref{fig:photocurrent}(b) shows the imprecision noise in the angular displacement obtained by implementing the calibration method.
The beam width at the position of the flat mirror is $\SI{587}{\micro m}$ measured by a beam profiler (BC207VIS, Thorlabs). 
The angular sensitivity improves as the optical power increases, and, at the maximum power of 5 mW, the imprecision noise reaches 2.56 $\t{prad}/\sqrt{\t{Hz}}$. 
This value is 16\% higher than the shot noise level (2.21 $\t{prad}/\sqrt{\t{Hz}}$) calculated under the assumption of ideal detection (\cref{PSD3}). Therefore, the detection efficiency is 0.75.

\subsection{Calibration of thermal noise spectra}

When measuring the torsion pendulum, we reduced the beam size from 587 µm to 180 µm to minimize clipping loss of the measurement beam at the finite surface width of the pendulum ($\SI{500}{\micro m}$). 
This reduction was achieved by focusing the beam onto the AOD at the input plane of the 4f system. 
However, it was observed that the AOD calibration method became inaccurate as the beam was focused onto the AOD crystal: despite applying a pure tone frequency modulation, the SPD signal was significantly distorted with harmonics. 
This appears to be due to the incoherent addition of diffraction angles as the beam is focused on the AOD crystal. 
To address this, we adopted a two-step calibration process: (1) calibrating the thermal motion of the torsion pendulum using a reliable, larger beam size ($w_0 \approx \SI{587}{\micro m}$), and (2) using this thermal motion as a reference to calibrate the SPD spectrum obtained with the smaller beam size ($w_0 \approx \SI{180}{\micro m}$).
The second step was performed with 1 mW optical power, which kept the pendulum at room 
temperature (290 K) (see \cref{sec:temp}). 
The calibration factor was determined by fitting the spectra to the thermal noise at room temperature, resulting in a calibration factor of $\alpha_{\t{cal}} = 1.65\times10^{-9}\, \t{rad/V}$ with a 2\% error.

Through two-step calibration, the imprecision noise at the SPD was found to be 
$S_\theta^{\t{imp}}=1.06\times10^{-22}$ at 10 mW optical power, corresponding to a detection efficiency of $0.244$.
The lower sensitivity is largely due to the reduced beam size (\cref{theta_imp}).

The ideal imprecision derived in (\cref{theta_imp}), which assumes an intact spatial mode profile upon reflection, does not hold for the torsion pendulum measurement: that derivation assumes that tilt is induced by an infinite
plane, whereas the finite extent of the torsion pendulum clips the wings of the Gaussian beam, leading to 
distortion of the reflected beam beyond the simple model. 
Nonetheless, the imprecision noise shown in Fig. 3(a) of the main text is quantum-noise-limited, as confirmed by its good agreement with the expected shot noise scaling (\cref{theta_imp}).

\section{Measurement-based feedback cooling}

\subsection{Theory}

The physical angular displacement of the torsion pendulum reacts to external torques, consisting of thermal, back-action, and feedback torques: 
\begin{equation}  
    \chi_0^{-1}[\Omega]\dop{\theta}_\t{phys}[\Omega]=\dop{\tau}_{\text{th}}[\Omega]+\dop{\tau}_{\text{ba}}[\Omega]
    +\hat{\tau}_{\text{fb}}[\Omega],
\end{equation}
where \(\chi_0[\Omega]\) is the susceptibility of the torsion pendulum. 
In measurement-based feedback control \cite{CourPin01,WilKip15}, the observed motion [\cref{thetaObs}]
\begin{equation}
    \dop{\theta}_{\t{obs}}[\Omega]=\dop{\theta}_\t{phys}[\Omega] + \dop{\theta}_{\text{imp}}[\Omega]
\end{equation}
is used to synthesize a feedback force, in this case a torque,
\begin{equation}
\dop{\tau}_{\t{fb}}[\Omega]=-\chi_{\t{fb}}^{-1}[\Omega]\dop{\theta}_{\t{obs}}[\Omega]
\end{equation}
so that the physical motion
\begin{equation}
    \dop{\theta}_\t{phys}[\Omega] = \chi_{\t{eff}}[\Omega](\dop{\tau}_\t{tot}[\Omega] 
    -\chi_{\t{fb}}^{–1}[\Omega]\dop{\theta}_{\t{imp}}[\Omega])
\end{equation}
is modified via the effective susceptibility
\(\chi_{\t{eff}}=(\chi_{\t{0}}^{-1}+\chi_{\t{fb}}^{-1})^{–1}\); 
here \(\dop{\tau}_{\t{tot}}=\dop{\tau}_{\t{th}}+\dop{\tau}_{\t{ba}}+\dop{\tau}_{\t{fb}}\).
In the presence of feedback, the observed motion is also modified:
\begin{equation}
    \dop{\theta}_{\t{obs}}[\Omega] = 
    \chi_{\t{eff}}[\Omega](\dop{\tau}_\t{tot}[\Omega]+\chi_{\t{0}}^{–1}[\Omega]
    \dop{\theta}_{\t{imp}}[\Omega])
\end{equation}

Cooling by this kind of feedback can be affected by the choice $-\chi_\t{fb} = i I \Omega \Gamma_\t{fb}$ around
mechanical resonance $\Omega_0$.
In this case, the spectrum of the observed motion assumes the form
\begin{equation}
S_\theta^{\text{obs}}[\Omega] = 
    \frac{S_{\tau}^{\text{tot}}[\Omega]}{I^2[(\Omega_{0}^2-\Omega^2)^2+(\Omega\Gamma_{\text{eff}})^2]}+\frac{(\Omega_{0}^2-\Omega^2)^2+(\Omega\Gamma_0[\Omega])^2}{(\Omega_{0}^2-\Omega^2)^2+(\Omega\Gamma_{\text{eff}})^2}S_{\theta}^{\text{imp}}[\Omega];
\end{equation}
this model is used to fit the data in Fig. 3 of the main text.
The spectrum of the physical motion
\begin{equation}\label{Sphys}
    S_\theta^\t{phys}[\Omega] = \frac{S_{\tau}^{\text{tot}}[\Omega]}{I^2[(\Omega_{0}^2-\Omega^2)^2+(\Omega\Gamma_{\text{eff}})^2]}+\frac{(\Omega\Gamma_{\text{fb}})^2}{(\Omega_{0}^2-\Omega^2)^2+(\Omega\Gamma_{\text{eff}})^2}S_{\theta}^{\text{imp}}[\Omega],
\end{equation}
allows inference of physical properties such as the average phonon occupation achieved by feedback cooling.

If we take the torsional motion to be a harmonic oscillator with creation/annihilation operators 
$\hat{b},\hat{b}^\dagger$, then the angular motion is $\hat{\theta} = \theta_\t{zp}(\hat{b}+\hat{b}^\dagger)$, 
where $\theta_\t{zp} = \sqrt{\hbar/(2I \Omega_0)}$. Then the variance of the motion is 
$\avg*{\hat{\theta}^2} = 2\theta_\t{zp}^2 (\avg*{\hat{b}^\dagger \hat{b}}+\tfrac{1}{2})$; thus the phonon occupation 
$n_\t{eff} = \avg*{\hat{b}^\dagger \hat{b}}$
can be inferred from the variance of the (physical) angular motion via
\begin{equation}
    n_\t{eff} + \frac{1}{2} = \frac{\avg*{\hat{\theta}^2}}{2\theta_\t{zp}^2} 
    = \int \frac{S_\theta^\t{phys}[\Omega]}{2\theta_\t{zp}^2} \frac{\dd \Omega}{2\pi}.
\end{equation}
Using the model in \cref{Sphys} gives
\begin{equation}
    n_\t{eff} + \frac{1}{2} = \left( n_\t{th}+n_\t{ba} + \frac{1}{2} \right)\frac{\Gamma_0}{\Gamma_\t{eff}} 
        + n_\t{imp}\frac{\Gamma_\t{eff}}{\Gamma_0},
\end{equation}
where $n_\t{ba} = S_\tau^\t{ba}[\Omega_0]/(4\hbar \Im \chi_0^{-1}[\Omega_0])$ is the effective back-action phonon occupation,
and $n_\t{imp} = S_\theta^\t{imp}/(2S_{\theta}^\t{zp}[\Omega_0])$ is the phonon-equivalent measurement imprecision.

\subsection{Implementation of feedback}

\begin{figure}[t!]
    \centering
    \includegraphics[width=0.9\columnwidth]{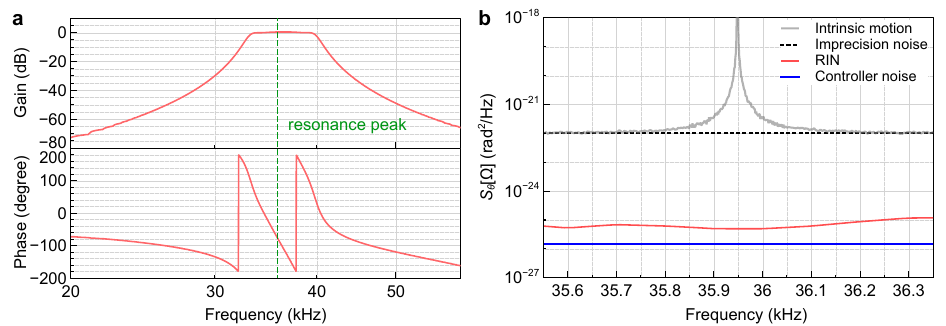}
    \caption{\label{fig:feedback}(a) Frequency response of the feedback controller. (b) Angular displacement-referred extraneous noises in the feedback loop.}
\end{figure}

The motion of the pendulum is actuated by radiation torque from a secondary laser beam (``actuation beam'')
focused on the edge of the pendulum’s backside.
A fiber-coupled electro-optic Mach-Zehnder amplitude modulator (LNX1020A, Thorlabs) is employed to 
modulate the radiation torque around the mechanical frequency.
The modulator’s response to the external voltage is sinusoidal due to the nature of the Mach-Zehnder 
interferometer in it; we bias the modulator around its linear operating point, which corresponds to the
bias voltage at which half the maximum optical power is transmitted.
At this operating point, the actuation beam exhibits a sensitivity of 15.7 mW/V in response to the external voltage
applied on the modulator.

\Cref{fig:feedback}(a) is the measured frequency response of the feedback filter $\chi_{\t{fb}}^{-1}$, optimized for feedback gain to achieve a phonon number of approximately 6000. This configuration is implemented using an FPGA-based controller (Moku:Pro, Liquid Instruments). The feedback filter comprises a bandpass filter (34–40 kHz) centered around the resonance frequency of the fundamental mode, and a phase delay that aligns the total loop delay to be $\pi/2$. As depicted in \cref{fig:feedback}(a), the delay at the resonance frequency $\Omega_0/2\pi=\SI{35.95}{kHz}$ is precisely set to $\pi/2$.

\subsection{Extraneous noise in feedback}

Extraneous noise in the feedback loop can limit the performance of feedback cooling if they are larger than the
imprecision noise in the measurement that drives the actuator.
We investigate this possibility by measuring and budgeting the noise in the actuator beam. 
\Cref{fig:feedback}(b) shows this budget referred to angular displacement at the SPD.
Two contributions can be distinguished: (a) voltage noise from the feedback controller, 
referred to angle (blue); and, (b) intensity noise in the actuation beam referred to angle (red). 
Grey shows the motional signal at the SPD, with black dashed showing the quantum-noise-limited imprecision.
Clearly, the extraneous noise in the feedback loop lies more than 30 dB below the imprecision noise.
Thus, feedback cooling is governed primarily by the observed motion.

\begin{figure}[t!]
    \centering
    \includegraphics[width=0.5\columnwidth]{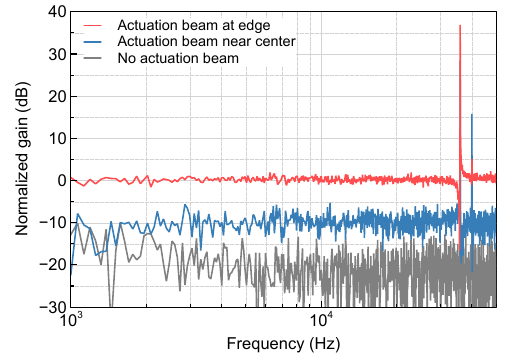}
    \caption{\label{fig:actuator} Frequency response of angular displacement of the torsion pendulum to the radiation pressure torque actuator.}
\end{figure}

Next, we investigate whether the actuation beam induces mechanical torques 
in excess of that due to radiation pressure. For example, thermoelastic torques \cite{MetKar04,KlecBouw06} 
arising from concentrated photothermal heating at the edge of the torsion pendulum. 
The nature of the thermoelastic effect can be understood through the absorption and diffusion of heat within 
the pendulum.
This suggests that the thermoelastic effect can be distinguished from pure radiation pressure 
torque \cite{SchKip06,Sudhir17}: the angular displacement response to thermoelastic torque follows a 
single-pole low-pass filter characteristic at a specific cutoff frequency, while the radiation pressure 
torque remains frequency-independent.

\Cref{fig:actuator} is the frequency-response of the torsion pendulum as the intensity in the actuator beam 
is modulated. 
The response remains frequency-independent from 1 kHz up to the resonance for various beam positions, 
both at the edge and near the center of the pendulum. 
This suggests that no thermoelastic effect is observed during feedback actuation, and that the actuation is
dominated by radiation torque.

\subsection{Estimation of mode temperature}\label{sec:temp}

To investigate the role of photothermal heating, we measured the mode temperature as a function of optical 
power (from now on, the optical power is referred to as that reflected from the torsion pendulum). For this, we fit the torsion pendulum’s intrinsic motion to a Lorentzian model: 
$S_{\theta}^{\t{obs}}[\Omega]=S_{\theta}^{\t{imp}}+S_{\theta}^{\t{th}}[\Omega_0]/(1+4Q_0^2(\Omega-\Omega_0)^2/\Omega_0^2)$. 
Since the resolution bandwidth of the spectrum (0.25 Hz) is larger than the linewidth of the torsional 
mode ($\Gamma_0/2\pi = \SI{2.6}{mHz}$), we excluded the vicinity of the peak from the fitting range.
\begin{figure}
    \centering
    \includegraphics[width=0.5\columnwidth]{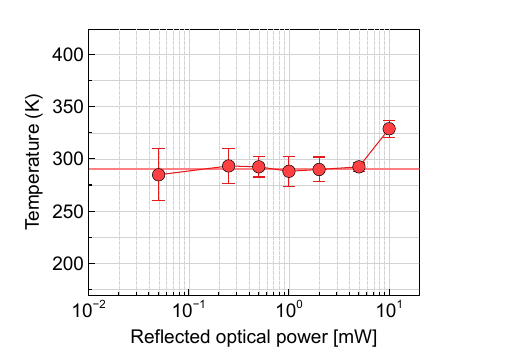}
    \caption{\label{fig:temp}
    Estimation of mode temperature using the AOD-based calibration method.
    }
\end{figure}
Using these fittings, we estimated the mode temperatures shown in \cref{fig:temp}. 
For optical powers below 5 mW, no signs of photothermal heating were observed.
The fits allow us to infer the effective moment of inertia $I = \SI{5.54e-17}{kg\cdot m^2}$ with the independently 
measured values of $\Omega_0=2\pi\cdot 35.95 \t{kHz}$, $Q=1.365\cdot10^7$, and $T = \SI{290}{K}$. 
This value is $30\%$ smaller than that obtained by finite element simulation 
(COMSOL Multiphysics), ($\SI{7.87e-17}{kg\cdot m^2}$) and  larger by $12\%$ compared to the analytical model
\cite{PraWil23} $I=\rho L h w^3/24 \approx \SI{4.91e-17}{kg\cdot m^2}$ of an ideal rectangular beam (which 
our device is not). This provides an independent consistency check of our angle calibration procedure and
mode temperature.

\end{document}